\documentclass{jpsj3}

\usepackage{bm}
\usepackage{amsmath}
\usepackage{amssymb}
\usepackage{array,booktabs}
\usepackage{graphicx}
\usepackage{float}

\title{Pairing mechanism of unconventional superconductivity in 
doped Kane-Mele model}

\author{Yuri Fukaya, Keiji Yada, Ayami Hattori, and Yukio Tanaka}
\inst{\address{Department of Applied Physics, Nagoya University, Chikusa, Nagoya 464-8603, Japan}}

\abst{We study the pairing symmetry of a doped Kane-Mele model on a honeycomb
lattice with on-site Coulomb 
interaction. 
The pairing instability of Cooper pair is calculated based on the 
linearized $\acute{\mathrm{E}}$liashberg equation within the random phase 
approximation (RPA). 
When the magnitude of the spin-orbit coupling is weak, 
even-frequency spin-singlet even-parity (ESE) pairing 
is dominant. 
On the other hand, with the increase of the spin-orbit coupling, 
we show that the even-frequency spin-triplet odd-parity (ETO) 
$f$-wave pairing exceeds ESE one. 
ETO $f$-wave pairing is 
supported by the longitudinal spin fluctuation. 
Since the transverse spin fluctuation is strongly 
suppressed by spin-orbit coupling, ETO $f$-wave pairing 
becomes dominant for large magnitude of spin-orbit coupling. 
}

\kword{Kane-mele model, unconventional superconductivity, 
spin-triplet pairing, Eliashberg equation, disconnected Fermi surface, 
honeycomb lattice}
\begin{document}
\maketitle
%
%
\section{Introduction}
To explore unconventional pairing in superconductivity has 
been an important issue in condensed matter physics\cite{Ueda}. 
It is known that $d$-wave pairing has been realized in 
high $T_{\mathrm{c}}$ cuprate \cite{Harlingen,Moriya,Nagaosa,Fukuyama} 
and there have been many remarkable 
quantum phenomena specific to unconventional pairing 
having sign changes of gap function 
on the Fermi surface \cite{Harlingen,Tsuei,Kashiwaya00}. 
There are several strongly correlated systems where 
spin-singlet $d$-wave pairing are realized. 
On the other hand, spin-triplet $p$-wave 
pairing is realized in superfluid 
$^{3}$He \cite{Leggett}. In solid state materials, pairing 
symmetry of Sr$_{2}$RuO$_{4}$ is believed to be 
spin-triplet $p$-wave pairing \cite{Maeno}. 

As a  natural extension of these anisotropic 
pairing, spin-triplet  $f$-wave pairing has been also 
proposed \cite{Kuroki2001,Tanaka2004,Kuroki2004,
Ikeda2004,Yanase2004,Nishikawa,Kuroki2005}.  
Since $f$-wave pairing has a higher angular momentum, 
its gap function must have sign changes much more as 
compared to $d$-wave and $p$-wave pairings. 
Thus, it cannot be stable 
due to the presence of many nodes on the Fermi surface 
as far as we are considering simple Fermi surface 
located around the $\Gamma$ point. 
However, as proposed by Kuroki $et.$ $al.$, 
$f$-wave pairing is possible if we 
consider disconnected Fermi surfaces since 
the gap nodes 
do not have to cross the  Fermi surface \cite{Kuroki2001}. 

One of the possible  systems is 
quasi one-dimensional organic superconductor 
(TMTSF)$_{2}$X (X=PF$_{6}$, ClO$_{4}$, etc. ). \cite{Jerome} 
A remarkable feature of this system is the 
coexistence of 2k$_{F}$ charge density wave 
and 2k$_{F}$ spin density wave (SDW). 
Then, the charge fluctuation becomes important and 
it favors the realization of spin-triplet pairing \cite{Tanaka2004,Kuroki2005}. 
Based on a fluctuation mediated pairing mechanism, 
$d$-wave and $f$-wave pairings become possible candidates.
There have been several theoretical studies which support 
realization of spin-triplet $f$-wave pairing 
\cite{Tanaka2004,Nickel,Fuseya,Kuroki2005,AizawaPRB,AizawaPRL,AizawaJPSJ}. 

Another possibility of 
$f$-wave pairing was intensively discussed just after the 
discovery of superconductivity in  $\mathrm{Na}_{x}$CoO$_{2}$ $\cdot$ $y$H$_{2}$O \cite{Takada}.
A triangular lattice structure of this material  
can host the disconnected Fermi surface around
the K and K' points. 
Spin-triplet $f$-wave pairing was 
proposed based on the fluctuation exchange method 
\cite{Kuroki2004,KurokiPRB2005}. 
Although there have been several theories supporting 
$f$-wave pairing \cite{Nishikawa,Ikeda2004,Yanase2004,Yanase2005,Yanase2005SO,Mazin,Mochizuki}, due to the 
presence of conflicting results\cite{Yada2,Yada1,Msato}, 
the pairing mechanism of this material is 
still controversial. 
 
Other than these materials, 
there are several unconventional superconductors, $e.g.$, 
UPt$_{3}$ \cite{Sauls,Tsutsumi} and SrPtAs \cite{Goryo,SrPtAs},
where the possibility of spin-triplet $f$-wave pairing 
has been suggested. 
Also, in optical lattice systems, spin-triplet $f$-wave pairing 
has been proposed \cite{DasSarma2010}. 
In the light of the preexisting theories, 
to explore spin-triplet $f$-wave pairing in 
hexagonal structures \cite{Honerkamp,ThomaleG,ThomaleK1,ThomaleK2,Chubkov} 
is a challenging issue. 

Recently, Zhang $et.$ $al.$ proposed that spin-triplet $f$-wave 
pairing is possible in doped silicene by 
applying an electric field \cite{Yao2015}. 
Silicene, single atomic layer of Si forming a 2D honeycomb lattice 
like graphene\cite{Novoselov666}, 
becomes a topical material from the 
view points of monolayer material and topological insulator. 
Nowadays, there are several works 
to explore unconventional superconductivity in 
atomic layered systems \cite{MoS2Iwasa,Chubkov2012,SiliceneSCgap,Law2014,Annica,Baskaran}. 
Thus,  
to study the superconductivity in doped Kane-Mele model is 
interesting since it is a canonical model of 
monolayer systems with non-trivial topological property\cite{KaneMelemodel1,KaneMelemodel2}. 
We naively expect that the spin-orbit coupling 
may  help the generation of spin-triplet pairing. 

In this paper, 
we study the pairing instability of Cooper pair in 
doped Kane-Mele model with on-site Coulomb interaction 
by the linearized $\acute{\mathrm{E}}$liashberg equation within the random phase 
approximation (RPA). 
We clarify that even-frequency spin-singlet even-parity 
(ESE) pairing is dominant 
when the magnitude of the spin-orbit coupling is weak. 
On the other hand, with the increase of the spin-orbit coupling, 
we show that even-frequency spin-triplet odd-parity (ETO) 
$f$-wave pairing becomes dominant. 
We clarify physical reasons why $f$-wave pairing is realized. 

The organization of this paper is as follows. 
In section II, we show a model Hamiltonian 
and formulations of the pairing interaction 
within RPA. An $\acute{\mathrm{E}}$liashberg equation is also formulated. 
In section III, we show calculated results  
of the $\acute{\mathrm{E}}$liashberg equation and 
discuss the pairing mechanism. 
In section IV, we summarize our results. 

\begin{figure}[H]
 \centering
 \includegraphics[width=5cm]{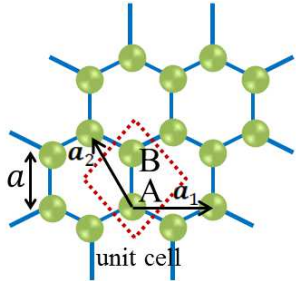}
 \caption{The structure of a honeycomb lattice. The dotted line denotes the unit cell where $\bm{a}_1$ and $\bm{a}_2$ are the lattice vectors.
Each unit cell contains two sublattice $A$ and $B$ with distance $a$.}\label{fig:honeycomb}
\end{figure}
%
\section{Model and Formulation}

\subsection{Hamiltonian}
\begin{figure}[htbp]
 \centering
  \includegraphics[width=12cm]{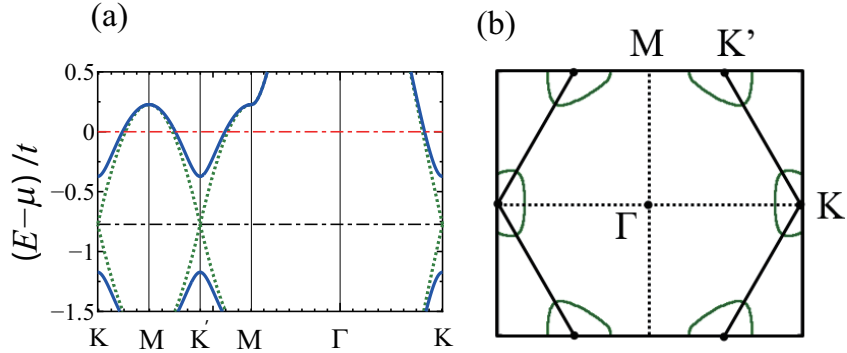}
  \caption{(a)The dispersion relation at $\lambda_\mathrm{SO}/t=$0 (dotted lines) and 0.4 (solid lines). $\mu/t=0.77$ for both cases.
  (b)Fermi surface in the normal states at $\lambda_\mathrm{SO}/t=$0.4 and 
  $\mu/t=0.77$. 
  The hexagonal line shows the first Brillouin zone.  }
\label{fig:dispersion}
\end{figure}

In this section, we introduce a model Hamiltonian and the formulations of the $\acute{\mathrm{E}}$liashberg equation to calculate the instabilities of the Cooper pairs.
We consider the honeycomb lattice as shown in Fig. \ref{fig:honeycomb}.
Here, we take lattice vectors as  ${\bm{a}}_{1}=(\sqrt{3}a,0)$ and ${\bm{a}_{2}}=(-\sqrt{3}a/2,3a/2)$.
On this lattice, we consider a Kane-Mele model\cite{KaneMelemodel1,KaneMelemodel2},
\begin{align}
&\mathcal{H}_{0}=\sum_{\bm{k}\sigma} \hat c^{\dag}_{\bm{k}\sigma}\hat H_{\sigma}(\bm{k}) \hat c_{\bm{k}\sigma},\label{eq:kane-mele}\\
&\hat H_{\sigma}(\bm{k})=\notag\\
&\begin{pmatrix}
-\mu+(\sigma_z)_{\sigma\sigma}\lambda_\mathrm{SO}W_\mathrm{SO}(\bm{k})&tW(\bm{k})\\
tW^*(\bm{k})&-\mu-(\sigma_z)_{\sigma\sigma}\lambda_\mathrm{SO}W_\mathrm{SO}(\bm{k})
\end{pmatrix},
\\
&W({\bm k})=(1 + e^{ -i \bm{k} \cdot {\bm{a}}_{2} } +e^{ -i \bm{k} \cdot ( {\bm{a}}_{1} + {\bm{a}}_{2} ) }),\\
&W_\mathrm{SO}({\bm k})=\frac{2 }{3 \sqrt{3}}\{\sin \bm{k} \cdot {\bm{a}}_{1} +\sin \bm{k} \cdot {\bm{a}}_2 -\sin \bm{k} \cdot ( {\bm{a}}_{1}+{\bm{a}}_{2})\},
\end{align}%
where $\hat c^\dag_{\bm{k}\sigma}=(c^\dag_{\bm{k}A\sigma}\ c^\dag_{\bm{k}B\sigma})$ and $\hat c_{\bm{k}\sigma}=(c_{\bm{k}A\sigma}\ c_{\bm{k}B\sigma})^T$ are creation and annihilation operators of the electron with momentum ${\bm k}$ and spin $\sigma$ ($\sigma=\uparrow$ or $\downarrow$) on sublattice $A$ and $B$. $\mu$, $t$ and $\lambda_\mathrm{SO}$ denote the chemical potential, the nearest-neighbor hopping and the intrinsic spin-orbit interaction, respectively. $\sigma_z$ is the Pauli matrix in spin space.
By diagonalizing $\hat H_{\sigma}(\bm{k})$, we obtain the dispersion relation in the normal state,
\begin{align}
 E_\sigma^\pm(\bm{k})=-\mu \pm \sqrt{t^2|W(\bm{k})|^{2}+\lambda_\mathrm{SO}^2W_\mathrm{SO}(\bm{k})^{2}}.
\end{align}
Since the spin-orbit interaction considered in the present model does not break the inversion symmetry and the time-reversal symmetry,
the energy bands are doubly degenerated.
In other words, $E_\sigma^\pm(\bm{k})$ does not depend on $\sigma$.
Without the spin-orbit interaction $\lambda_\mathrm{SO}$, 
the valence bands and the conduction bands touch at the K and K' points because $W(\bm{k})=0$ there.
However, $\lambda_\mathrm{SO}$ makes  band gaps as shown in Fig. \ref{fig:dispersion}.
To study the superconductivity in this system, we consider slightly 
carrier doped metallic state in the conduction bands.
In this paper, we choose $\mu$ as the number 
of carrier in the conduction bands becomes 0.1 in each spin component. 
The obtained Fermi surfaces at $\lambda_\mathrm{SO}/t=0.4$ are shown in 
Fig. \ref{fig:dispersion}(b).
As seen in the figure, two disconnected electron-pockets are formed at 
the K and K' points.
The shape of the Fermi surfaces is almost the same for other values of 
$\lambda_\mathrm{SO}$ used in the present paper, 
and the above character does not change.

As well as the non-interacting term in Eq. (\ref{eq:kane-mele}), 
we introduce the on-site repulsive interaction,
\begin{align}
  \mathcal{H}_{I}&=\frac{U}{N}\sum_{{\bm k}{\bm k}'{\bm q}\alpha}c^{\dagger}_{\bm{k+q}\alpha\uparrow}c^{\dagger}_{\bm{k'-q}\alpha\downarrow}c_{\bm{k'}\alpha\downarrow} c_{\bm{k}\alpha\uparrow}.
\end{align}
Here, $U$ and $N$ represent the on-site repulsive interaction and the system size. 
This interaction is treated by RPA.

\subsection{Susceptibilities and effective pairing interactions}
In this subsection, we calculate susceptibilities and resulting pairing interactions in the framework of RPA.
For this purpose, we introduce the non-interacting temperature Green's function,
\begin{align}
\hat{G}^{\sigma}(\bm{k},i\varepsilon_n)&=(i \varepsilon_n - H_{\sigma}(\bm{k}))^{-1}\notag\\
&=
\begin{pmatrix}
G_{AA}^{\sigma}(\bm{k},i \varepsilon_n) & G_{AB}^{\sigma}(\bm{k},i \varepsilon_n)\\
G_{BA}^{\sigma}(\bm{k},i \varepsilon_n) & G_{BB}^{\sigma}(\bm{k},i \varepsilon_n)
\end{pmatrix},
\end{align}
where $\varepsilon_n=(2n+1)\pi k_\mathrm{B}T$ is a fermionic Matsubara frequency.
Then, the irreducible susceptibilities are given by
\begin{align}
 \chi_{\alpha\beta;\gamma\delta}^{\sigma\tau}(\bm{q}, i\omega_{m}) 
 &=-\frac{k_\mathrm{B}T}{N} \sum_{\bm{k}, i\varepsilon_n}
G_{\alpha\gamma}^\sigma(\bm{k}+\bm{q},i\varepsilon_{n}+i\omega_{m})
G_{\delta\beta}^{\tau}(\bm{k},i\varepsilon_n),
\end{align}
where $\omega_m=2m\pi k_\mathrm{B}T$ is a bosonic Matsubara frequency. 
$\alpha$, $\beta$, $\gamma$ and $\delta$ ($\sigma$ and $\tau$) 
indicate sublattice (spin) indeces. 
From these irreducible susceptibilities, 
we construct bubble and ladder-type diagrams to calculate the spin and charge susceptibilities,
\begin{align}
&\chi_{\alpha\sigma,\beta\tau}^\mathrm{B}(\bm{q}, i\omega_{m})\notag\\
&=\delta_{\sigma\tau} \chi_{\alpha\alpha;\beta\beta}^{\sigma\sigma}(\bm{q}, i\omega_{m})
-\sum_{\gamma}\chi_{\alpha\alpha;\gamma\gamma}^{\sigma\sigma}(\bm{q}, i\omega_{m})U\chi_{\gamma\bar{\sigma},\beta\tau}^\mathrm{B}(\bm{q}, i\omega_{m}),\label{eq:bubble}\\
&\chi_{\alpha\sigma,\beta\sigma}^\mathrm{L}(\bm{q}, i\omega_{m})\notag\\
&=\chi_{\alpha\alpha;\beta\beta}^{\sigma\bar{\sigma}}(\bm{q}, i\omega_{m})
+\sum_{\gamma}\chi_{\alpha\alpha;\gamma\gamma}^{\sigma\bar{\sigma}}(\bm{q}, i\omega_{m})U
\chi_{\gamma\sigma,\beta\sigma}^\mathrm{L}(\bm{q}, i\omega_{m}),\label{eq:laddar}
\end{align}
where $\bar{\sigma}=\uparrow$ and $\downarrow$ for $\sigma=\downarrow$ and $\uparrow$, respectively.
Note that $\chi_{\alpha\sigma,\beta\bar{\sigma}}^\mathrm{L}(\bm{q}, i\omega_{m})$ with $\sigma\neq\bar{\sigma}$ are absent since there is no spin-flipping term in the non-perturbative Hamiltonian and on-site interaction acts between electrons with opposite spins.
By solving simultaneous equations in Eqs. (\ref{eq:bubble}) and (\ref{eq:laddar}), we obtain $\chi_{\alpha\sigma,\beta\tau}^\mathrm{B}(\bm{q}, i\omega_{m})$ and $\chi_{\alpha\sigma,\beta\sigma}^\mathrm{L}(\bm{q}, i\omega_{m})$ as,
\begin{align}
&\chi_{\alpha\sigma,\alpha\sigma}^\mathrm{B}=(\chi_{\alpha\sigma,\alpha\sigma}^\mathrm{B0}+U^2\chi_{\bar{\alpha}\bar{\sigma},\bar{\alpha}\bar{\sigma}}^\mathrm{B0}\phi^\mathrm{B0}_\sigma)/D^\mathrm{B},\\
&\chi_{\alpha\sigma,\bar{\alpha}\sigma}^\mathrm{B}=(\chi_{\alpha\sigma,\bar{\alpha}\sigma}^\mathrm{B0}-U^2\chi_{\alpha\bar{\sigma},\bar{\alpha}\bar{\sigma}}^\mathrm{B0}\phi^\mathrm{B0}_\sigma)/D^\mathrm{B},\\
&\chi_{\alpha\sigma,\alpha\bar{\sigma}}^\mathrm{B}=\left(-U\sum_\beta\chi_{\alpha\sigma,\beta\sigma}^\mathrm{B0}\chi_{\beta\bar{\sigma},\alpha\bar{\sigma}}^\mathrm{B0}+U^3\Phi^\mathrm{B0}\right)/D^\mathrm{B},\\
&\chi_{\alpha\sigma,\bar{\alpha}\bar{\sigma}}^\mathrm{B}=\left(-U\sum_\beta\chi_{\alpha\sigma,\beta\sigma}^\mathrm{B0}\chi_{\beta\bar{\sigma},\bar{\alpha}\bar{\sigma}}^\mathrm{B0}\right)/D^\mathrm{B},\\
\end{align}
with
\begin{align}
\chi_{\alpha\sigma,\beta\sigma}^\mathrm{B0}&=\chi_{\alpha\alpha;\beta\beta}^{\sigma\sigma},\\
D^\mathrm{B}&=1-U^2\sum_{\alpha\beta}\chi_{\alpha\uparrow,\beta\uparrow}^\mathrm{B0}\chi_{\beta\downarrow,\alpha\downarrow}^\mathrm{B0}+U^4\Phi^\mathrm{B0},\\
\phi^\mathrm{B0}_\sigma&=\chi_{A\sigma,B\sigma}^\mathrm{B0}\chi_{B\sigma,A\sigma}^\mathrm{B0}-\chi_{A\sigma,A\sigma}^\mathrm{B0}\chi_{B\sigma,B\sigma}^\mathrm{B0},\\
\Phi^\mathrm{B0}&=\phi^\mathrm{B0}_\uparrow\phi^\mathrm{B0}_\downarrow,
\end{align}
and
\begin{align}
&\chi_{\alpha\sigma,\alpha\sigma}^\mathrm{L}=(\chi_{\alpha\sigma,\alpha\sigma}^\mathrm{L0}+U\phi^\mathrm{L0}_\sigma)/D^\mathrm{L}_\sigma,\\
&\chi_{\alpha\sigma,\bar{\alpha}\sigma}^\mathrm{L}=\chi_{\alpha\sigma,\bar{\alpha}\sigma}^\mathrm{L0}/D^\mathrm{L}_\sigma,
\end{align}
with
\begin{align}
\chi_{\alpha\sigma,\beta\sigma}^\mathrm{L0}&=\chi_{\alpha\alpha;\beta\beta}^{\sigma\bar{\sigma}},\\
\phi^\mathrm{L0}_\sigma&=\chi_{A\sigma,B\sigma}^\mathrm{L0}\chi_{B\sigma,A\sigma}^\mathrm{L0}-\chi_{A\sigma,A\sigma}^\mathrm{L0}\chi_{B\sigma,B\sigma}^\mathrm{L0},\\
D^\mathrm{L}_\sigma&=1-U\sum_{\alpha}\chi_{\alpha\sigma,\alpha\sigma}^\mathrm{L0}-U^2\phi^\mathrm{L0}_\sigma,
\end{align}
where we abbreviate the variable $\bm{q}$ and $i\omega_m$.
Then, we derive the longitudinal and transverse spin and charge susceptibilities,
\begin{align}
& \chi^{zz}_{\alpha\beta}(\bm{q}, i\omega_{m})
=\frac{1}{4}\sum_{\sigma}(\chi_{\alpha\sigma,\beta\sigma}^\mathrm{B}(\bm{q},i\omega_{m})-\chi_{\alpha\sigma,\beta\bar{\sigma}}^\mathrm{B}(\bm{q},i\omega_{m})),\\
& \chi^{+-}_{\alpha\beta}(\bm{q}, i\omega_{m})
=\chi_{\alpha\sigma,\beta\sigma}^\mathrm{L}(\bm{q},i\omega_{m}),\\
&\chi^\mathrm{C}_{\alpha\beta}(\bm{q}, i\omega_{m})
=\frac{1}{2}\sum_{\sigma\sigma'}\chi_{\alpha\sigma,\beta\sigma'}^\mathrm{B}(\bm{q},i\omega_{m}),
\end{align}
where  $\chi^{zz}_{\alpha\beta}(\bm{q}, i\omega_{m})$, 
$\chi^{+-}_{\alpha\beta}(\bm{q}, i\omega_{m})$ and
$\chi^\mathrm{C}_{\alpha\beta}(\bm{q}, i\omega_{m})$ 
denote 
longitudinal spin, transverse spin 
and charge susceptibilities, respectively.
Without the spin-orbit interaction, 
spin rotational symmetry leads to the relation $\chi^{+-}_{\alpha\beta}(\bm{q}, i\omega_{m})=2\chi^{zz}_{\alpha\beta}(\bm{q}, i\omega_{m})$.
However, this relation is broken in the presence of spin-orbit interaction.
When $D^\mathrm{B}$ ($D^\mathrm{L}_\sigma$) becomes 0, longitudinal (transverse) spin susceptibility diverges.
In other words, we can determine the critical temperature for magnetic instability by solving these equations.

\subsection{Linearized $\acute{E}$liashberg equation}
In this subsection, we introduce the $\acute{\mathrm{E}}$liashberg equation to discuss the pairing instability.
The $\acute{\mathrm{E}}$liashberg equation is used to determine the critical temperature $T_{\mathrm{c}}$ of superconductivity and gap function just below $T_{\mathrm{c}}$.
In this temperature region, the gap function and the anomalous Green's function can be linearized in the Dyson-Gor'kov equation.
Then, the Dyson-Gor'kov equation can be reduced to the eigenvalue equation.
This eigenvalue equation is called the $\acute{\mathrm{E}}$liashberg equation given by
\begin{align}
 &\lambda \Delta_{\alpha \sigma, \beta \tau}(\bm{k}, i\varepsilon_n) \notag \\
       &= -\frac{k_BT}{N} \sum_{\bm{k}', i\varepsilon_{n'}} \sum_{\gamma , \delta} V_{\alpha \sigma,\beta \tau;\gamma \sigma',\delta \tau'}(\bm{k}-\bm{k}',i\varepsilon_n - i\varepsilon_{n'}) F_{\gamma \sigma', \delta \tau'}(\bm{k}' , i\varepsilon_{n'}),\\
 &F_{\gamma \sigma', \delta \tau'}(\bm{k}',i\varepsilon_{n'}) \notag \\
 &= \sum_{\alpha' ,\beta'} G_{\gamma\alpha'}^{\sigma'}(\bm{k}' , i\varepsilon_{n'}) \Delta_{\alpha' \sigma', \beta' \tau'}(\bm{k}' , i\varepsilon_{n'} ) G_{\delta\beta'}^{\tau'}(-\bm{k}' , -i\varepsilon_{n'}),
\end{align}
where $\lambda$ denotes the eigenvalue. $V_{\alpha \sigma,\beta \tau;\gamma \sigma',\delta \tau'}(\bm{q},i\omega_m)$,  $\Delta_{\alpha \sigma, \beta \tau}(\bm{k}, i\varepsilon_n)$, and 
$F_{\gamma \sigma', \delta \tau'}(\bm{k}' , i\varepsilon_{n'})$ are 
effective pairing interaction, energy gap function and anomalous Green's function, respectively. 
Effective pairing interactions 
are given by 
\begin{align}
V_{\alpha\sigma,\beta\bar{\sigma};\alpha\sigma,\beta \bar{\sigma}}(\bm{q},i\omega_m)&=U\delta_{\alpha\beta}-U^2\chi_{\alpha\bar{\sigma},\beta\sigma}^\mathrm{B}(\bm{q},i\omega_{m}),\\
V_{\alpha\sigma,\beta\bar{\sigma};\alpha\bar{\sigma},\beta\sigma }(\bm{q},i\omega_m)&=-U^2\chi_{\alpha\sigma,\beta\sigma}^\mathrm{L}(\bm{q},i\omega_{m}),\\
V_{\alpha\sigma,\beta\sigma;\alpha\sigma,\beta\sigma }(\bm{q},i\omega_m)&=-U^2\chi_{\alpha\bar{\sigma},\beta\bar{\sigma}}^\mathrm{B}(\bm{q},i\omega_{m}).
\end{align}
Using these pairing interactions and the property of Fermi-Dirac statistics, $i.e.$,  $\Delta_{\alpha \sigma, \beta \tau}(\bm{k}, i\varepsilon_n)=-\Delta_{\beta \tau,\alpha \sigma}(-\bm{k}, -i\varepsilon_n)$, we obtain the $\acute{\mathrm{E}}$liashberg equations for $(\sigma, \tau)=(\uparrow,\uparrow)$, 
$(\uparrow,\downarrow)$, $(\downarrow,\uparrow)$ 
and $(\downarrow,\downarrow)$. 
In the present system, inversion symmetry exists in 
the normal state.
Thus, the solutions of the $\acute{\mathrm{E}}$liashberg equation should be the eigenstate of the parity. 
There are ESE and odd-frequency spin-triplet 
even-parity (OTE) pairings in the even-parity states 
while  ETO and odd-frequency spin-singlet odd-parity (OSO) pairings 
in the odd-parity states. 
In general, the solutions of the $\acute{\mathrm{E}}$liashberg equation 
are mixture of even-frequency and odd-frequency pairings. 
Without $\lambda_{\mathrm{SO}}$, numerically obtained 
pairing symmetry is classified into 
i)ETO with $S_{z}=0$, 
ii)ETO with $S_{z}=1$, 
iii)ETO with $S_{z}=-1$, 
and 
iv)ESE  as shown in Table I. 
Former three pairings are degenerate due to the spin-rotational 
symmetry. 
This degeneracy is lifted by $\lambda_{\mathrm{SO}}$.  
However, the  spin-rotational symmetry  around 
the $z$-direction keeps the degeneracy of 
pairings ii) and iii).
In the presence of $\lambda_{\mathrm{SO}}$, 
OSO and OTE pairings become subdominant component 
of ETO pairing with $S_{z}=0$ and ESE one, respectively.   
There are no subdominant odd-frequency pairing in
ETO with $S_{z}=\pm 1$
since $\lambda_{\mathrm{SO}}$ preserves $S_{z}$.\par
In the $\acute{\mathrm{E}}$liashberg equation, $\lambda$ becomes unity at $T=T_{\mathrm{c}}$ and $\lambda$ increases with decreasing $T$.
Therefore, it is presumable that the eigenstate with maximum eigenvalue is 
the most stable pairing.
We find them by the power iteration method.
\begin{table}[htb]
 \centering
  \begin{tabular}{|c|c|} \hline
    pairing symmetry  & induced odd-frequency \\
    without $\lambda_{\mathrm{SO}}$ & pairing by $\lambda_{\mathrm{SO}}$ \\ \hline \hline
    ETO$(\uparrow\downarrow+\downarrow\uparrow)$ & OSO$(\uparrow\downarrow-\downarrow\uparrow)$  \\ \hline
    ETO$(\uparrow\uparrow)$ & no \\ \hline
    ETO$(\downarrow\downarrow)$ & no \\ \hline
    ESE$(\uparrow\downarrow-\downarrow\uparrow)$ & OTE$(\uparrow\downarrow+\downarrow\uparrow)$  \\ \hline
  \end{tabular}
\caption{Mixture of even-frequency and odd-frequency pairing by spin-orbit 
coupling }
\end{table}
\section{Results}
In the following, we fix temperature 
$k_{\mathrm{B}} T/t =0.04$, 
where $t$ is the hopping parameter
of the nearest neighbors. 
The system size $N$ and cut-off Matsubara frequency $\varepsilon_{nmax}$
are chosen as $N=64 \times 64$ and $nmax=2048$ 
to guarantee the numerical accuracy.
Before we show the calculated energy gap functions,  
we discuss the general properties about the 
symmetry of gap functions.
The spatial inversion operation 
changes the sign of $\bm{k}$ and exchanges the 
site indexes $A$ and $B$. 
Then,  
$\Delta_{\alpha \sigma, \beta \tau}(\bm{k},i\varepsilon_{n})
=\Delta_{\bar{\alpha} \sigma, \bar{\beta} \tau}(-\bm{k},i\varepsilon_{n})$ 
is satisfied for the even-parity pairing. 
Here, $\bar{\alpha}$ and $\bar{\beta}$ are taken as $\bar{\alpha} \neq \alpha$ 
and $\bar{\beta} \neq \beta$, respectively. 
Similarly, 
$\Delta_{\alpha \sigma, \beta \tau}(\bm{k},i\varepsilon_{n})
=-\Delta_{\bar{\alpha} \sigma, \bar{\beta} \tau}(-\bm{k},i\varepsilon_{n})$ 
is satisfied for the odd-parity pairing. 
In the case of even-parity pairing 
$\Delta_{\alpha\sigma,\beta\bar{\sigma}}(\bm{k},i\varepsilon_{n})$ 
is decomposed into 
ESE pairing and OTE pairing 
as follows, 
\begin{equation}
\Delta_{\alpha\sigma,\beta\bar{\sigma}}(\bm{k},i\varepsilon_{n})
=
\Delta^{\mathrm{ESE}}_{\alpha\sigma,\beta\bar{\sigma}}(\bm{k},i\varepsilon_{n})
+ \Delta^{\mathrm{OTE}}_{\alpha\sigma,\beta\bar{\sigma}}(\bm{k},i\varepsilon_{n}).
\end{equation}
$\Delta^{\mathrm{ESE}}_{\alpha\sigma,\beta\bar{\sigma}}
(\bm{k},i\varepsilon_{n})$
and 
$\Delta^{\mathrm{OTE}}_{\alpha\sigma,\beta\bar{\sigma}}(\bm{k},i\varepsilon_{n})$
have following relations, 
\begin{equation}
\Delta^{\mathrm{ESE}}_{\alpha\sigma,\beta\bar{\sigma}}
(\bm{k},i\varepsilon_{n})
=\Delta^{\mathrm{ESE}}_{\alpha\sigma,\beta\bar{\sigma}}
(\bm{k},-i\varepsilon_{n}),
\end{equation}
and 
\begin{equation}
\Delta^{\mathrm{OTE}}_{\alpha\sigma,\beta\bar{\sigma}}
(\bm{k},i\varepsilon_{n})
=-\Delta^{\mathrm{OTE}}_{\alpha\sigma,\beta\bar{\sigma}}
(\bm{k},-i\varepsilon_{n}),
\end{equation}%
respectively. 
\par

\begin{figure}[htb]
 \centering
 \includegraphics[width=12cm,clip]{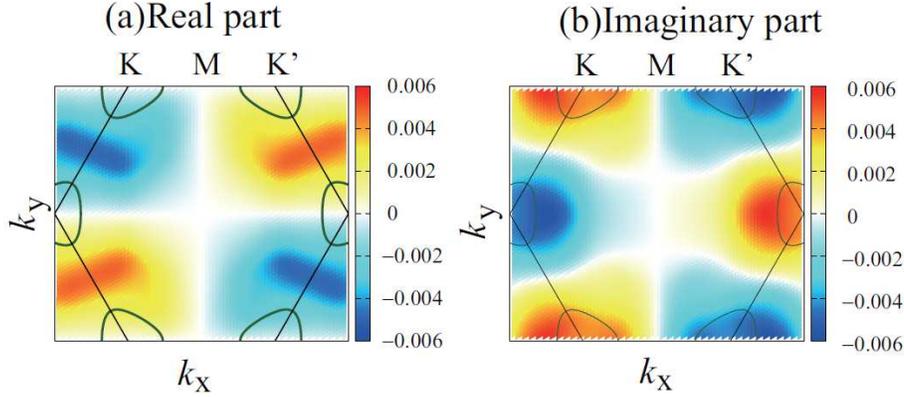}
 \caption{
 $\Delta_{A\uparrow,B\downarrow}(\bm{k},i\pi T)$
for ESE pairing with 
$\lambda_{\mathrm{SO}}/t=0.1$ and $U/t=2.75$: (a) Real part and 
(b)Imaginary part.
Real part is equivalent  to 
$\frac{1}{2}[\Delta_{A\uparrow,B\downarrow}(\bm{k},i\pi T)
+ \Delta_{B\uparrow,A\downarrow}(\bm{k},i\pi T)]$
and imaginary part is equivalent to 
$\frac{1}{2i}[\Delta_{A\uparrow,B\downarrow}(\bm{k},i\pi T)
- \Delta_{B\uparrow,A\downarrow}(\bm{k},i\pi T)]$
}
\label{ESE}
\end{figure}

First, we focus on the situation where spin-orbit coupling is 
not strong. 
The most dominant pairing is shown in 
Fig. \ref{ESE} 
where ESE pairing 
is realized. 
The obtained results are complicated owing to the 
honeycomb lattice structures including $A$ and $B$ sites. 
In the present choice of the gauge, the real part of 
$\Delta^{\mathrm{ESE}}_{A\uparrow,B\downarrow}(\bm{k},i\pi T)$
is an even-function of $\bm{k}$ and 
its imaginary part is an odd-function of 
$\bm{k}$. 
\begin{figure}[htb]
 \centering
 \includegraphics[width=10cm,clip]{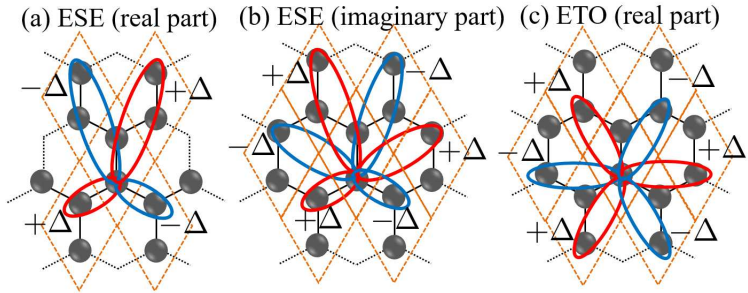}
 \caption{Schematic illustration of 
(a)Real part of ESE pairing, 
(b)Imaginary part of ESE pairing, and 
(c)Real part of ETO  $f$-wave pairing. 
The dashed lines denote the unit cell.
In case (c), imaginary part is negligible.  }
\label{illustration}
\end{figure}
The real part is interpreted as  a $d$-wave pairing and  
the corresponding imaginary part is $f$-wave like pairing as shown in 
Figs. \ref{illustration}(a) and (b). 
We have checked that 
${\mathrm{Re}}[\Delta^{\mathrm{ESE}}_{A\uparrow,B\downarrow}(\bm{k},i\pi T)]
={\mathrm{Re}}[\Delta^{\mathrm{ESE}}_{B\uparrow,A\downarrow}(\bm{k},i\pi T)]$ 
and 
${\mathrm{Im}}[\Delta^{\mathrm{ESE}}_{A\uparrow,B\downarrow}(\bm{k},i\pi T)]
=-{\mathrm{Im}}[\Delta^{\mathrm{ESE}}_{B\uparrow,A\downarrow}(\bm{k},i\pi T)]$
are kept.  
Then, 
$\Delta^{\mathrm{ESE}}_{A\uparrow,B\downarrow}(\bm{k},i\pi T)
=\Delta^{\mathrm{ESE}}_{B\uparrow,A\downarrow}(-\bm{k},i\pi T)$ 
is satisfied and this relation is consistent with even-parity  pairing. 
On the Fermi surface, the $f$-wave like imaginary component 
${\mathrm{Im}}[\Delta^{\mathrm{ESE}}_{A\uparrow,B\downarrow}(\bm{k},i\pi T)]$
is larger than the $d$-wave like real one 
${\mathrm{Re}}[\Delta^{\mathrm{ESE}}_{A\uparrow,B\downarrow}(\bm{k},i\pi T)]$. 
\par
\begin{figure}[htb]
 \centering
 \includegraphics[width=12cm,clip]{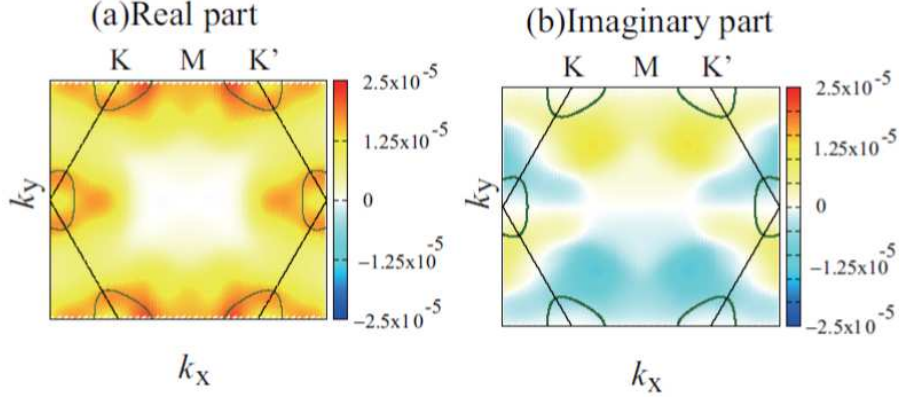}
 \caption{The subdominant OTE pairing 
$\Delta_{A\uparrow,B\downarrow}(\bm{k},i\pi T)$
for $\lambda_{\mathrm{SO}}/t=0.1$ and $U/t=2.75$: 
(a)Real part and  (b)Imaginary part. 
 }
\label{OTE}
\end{figure}

Besides this ESE pairing, there is a subdominant odd-frequency pairing 
which is almost two orders smaller than primary ESE pairing. 
As shown in Fig. \ref{OTE}, OTE 
pairing is induced. 
Similar to the primary ESE pairing, 
this induced odd-frequency gap function satisfies 
${\mathrm{Re}}[\Delta^{\mathrm{OTE}}_{A\uparrow,B\downarrow}(\bm{k},i\pi T)]
={\mathrm{Re}}[\Delta^{\mathrm{OTE}}_{A\uparrow,B\downarrow}(-\bm{k},i\pi T)]$ 
and 
${\mathrm{Re}}[\Delta^{\mathrm{OTE}}_{A\uparrow,B\downarrow}(\bm{k},i\pi T)]
={\mathrm{Re}}[\Delta^{\mathrm{OTE}}_{B\uparrow,A\downarrow}(\bm{k},i\pi T)]$ 
for real part, and 
${\mathrm{Im}}[\Delta^{\mathrm{OTE}}_{A\uparrow,B\downarrow}(\bm{k},i\pi T)]
=-{\mathrm{Im}}[\Delta^{\mathrm{OTE}}_{A\uparrow,B\downarrow}(-\bm{k},i\pi T)]$
and 
${\mathrm{Im}}[\Delta^{\mathrm{OTE}}_{A\uparrow,B\downarrow}(\bm{k},i\pi T)]
=-{\mathrm{Im}}[\Delta^{\mathrm{OTE}}_{B\uparrow,A\downarrow}(\bm{k},i\pi T)]$ 
for imaginary part. 
Then, following relation 
$\Delta^{\mathrm{OTE}}_{A\uparrow,B\downarrow}(\bm{k},i\pi T)
=\Delta^{\mathrm{OTE}}_{B\uparrow,A\downarrow}(-\bm{k},i\pi T)$ 
is satisfied to be consistent with even-parity. 

The reason of the generation of this 
subdominant OTE pairing is as follows.
First, we are taking into account the 
Matsubara frequency dependence 
of the effective pairing interactions 
in the process of solving the $\acute{\mathrm{E}}$liahsberg equation, 
then the existence of the odd-frequency pairing is allowed. 
Second, spin-orbit coupling breaks the 
spin-rotational symmetry, then it causes the mixture of spin-triplet component. 
Since we are considering 
intrinsic spin-orbit coupling without 
momentum dependence which does not 
flip spin, spin-triplet component with  
$S_{z}=0$ is mixed as a subdominant component. 
The solution in Figs. \ref{ESE} and \ref{OTE} belongs to the 
E$_{g}$ representation with double degeneracy. 
Then, 
the actual gap function might be a linear combination of 
these two solutions such as $d + id$-wave pairing. \par
With the increase of 
$\lambda_{\mathrm{SO}}$, the obtained pairing symmetry changes from 
ESE to ETO. 
As shown in Table. I, three kinds of spin state exist as a solution of 
the $\acute{\mathrm{E}}$liashberg equation. 
In the presence of $\lambda_{\mathrm{SO}}$, 
the degeneracy is lifted while that between $\uparrow\uparrow$ and 
$\downarrow\downarrow$ is kept. 
In the present calculation,  $\uparrow\uparrow$ and $\downarrow\downarrow$ 
spin states are more stabilized than $\uparrow\downarrow+\downarrow\uparrow$ one. 
\begin{figure}[htb]
 \centering
 \includegraphics[width=6cm,clip]{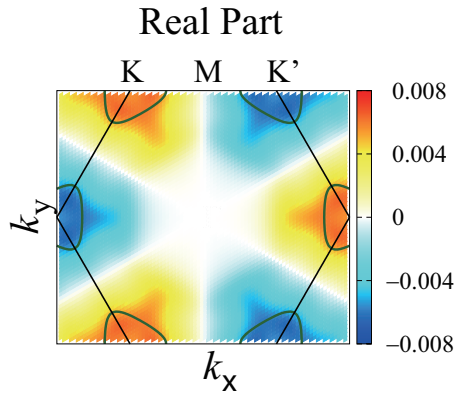}
 \caption{Real part of 
$\Delta_{A\uparrow,A\uparrow}(\bm{k},i\pi T)$
ETO pairing is plotted for 
$\lambda_{\mathrm{SO}}/t=0.5$, and $U/t=2.65$.
}
\label{ETO}
\end{figure}%
As shown in Fig. \ref{ETO}, the obtained gap function has a 
six-fold symmetry as a function of $\bm{k}$ and it 
is regarded as a $f$-wave pairing as shown in 
Fig. \ref{illustration}(c). 
In this case, the imaginary part of 
$\Delta_{A\uparrow,A\uparrow}(\bm{k},i\pi T)$ 
is negligible small on the Fermi surface. 
We have checked that 
$\Delta_{A\uparrow,A\uparrow}(\bm{k},i\pi T)
=-\Delta_{A\uparrow,A\uparrow}(-\bm{k},i\pi T)
$
and 
$\Delta_{A\uparrow,A\uparrow}(\bm{k},i\pi T)
=\Delta_{B\uparrow,B\uparrow}(\bm{k},i\pi T)
$. 
In this case, subdominant odd-frequency pairing 
never appears. 
This is because the presence of the spatial inversion 
symmetry. Since the present pair is odd-parity pairing, 
possible subdominant odd-frequency pairing has an 
OSO symmetry. 
On the other hand, there is no spin flipping term 
in the Hamiltonian.  
Owing to the parallel spin structure of 
dominant ETO pair, spin-singlet pair is 
prohibited. 
\par

\begin{figure}[htb]
 \centering
 \includegraphics[width=6cm,clip]{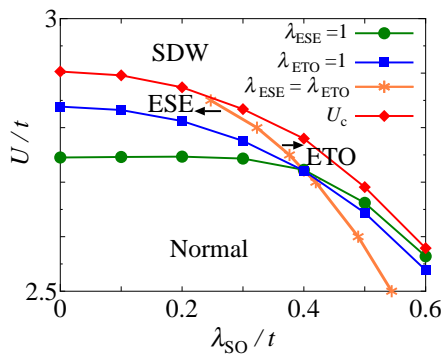}
 \caption{Phase diagram in the space of $U$ and $\lambda_{\mathrm{SO}}$ 
at $k_{\mathrm{B}}T/t=0.04$. 
$U_{\mathrm{c}}$ denotes the line above which longitudinal spin susceptibility 
diverges. $\lambda_{\mathrm{ESE}}=1$ ($\lambda_{\mathrm{ETO}}=1$) is a line where 
eigenvalue of  ESE pairing (ETO $f$-wave like pairing) 
diverges. 
}
\label{phase}
\end{figure}
In Fig. \ref{phase}, we show the phase diagram obtained 
in this model at $k_{\mathrm{B}}T/t=0.04$. 
With the increase of the on-site Coulomb interaction $U$, 
the longitudinal spin susceptibility diverges at   
$U=U_{\mathrm{c}}$, and SDW phase appears for 
$U>U_{\mathrm{c}}$. 
Above the line of $\lambda_{\mathrm{ESE}}=1$ 
($\lambda_{\mathrm{ETO}}=1$), ESE (ETO) pairing is stabilized.  
The pairing instability occurs due to the enhancement of the
spin-fluctuation near the SDW phase. 
The line connecting crossed mark shows the 
line of $\lambda_{\mathrm{ESE}}=\lambda_{\mathrm{ETO}}$ in this phase diagram. 
At the left (right) side of this line, 
eigenvalue $\lambda_{\mathrm{ESE}}$ is larger (smaller) than $\lambda_{\mathrm{ETO}}$. 
The interesting nature is the 
transition  from ESE pairing to 
ETO pairing with the increase of $\lambda_{\mathrm{SO}}$. 

\begin{figure}[htb]
 \centering
 \includegraphics[width=12cm,clip]{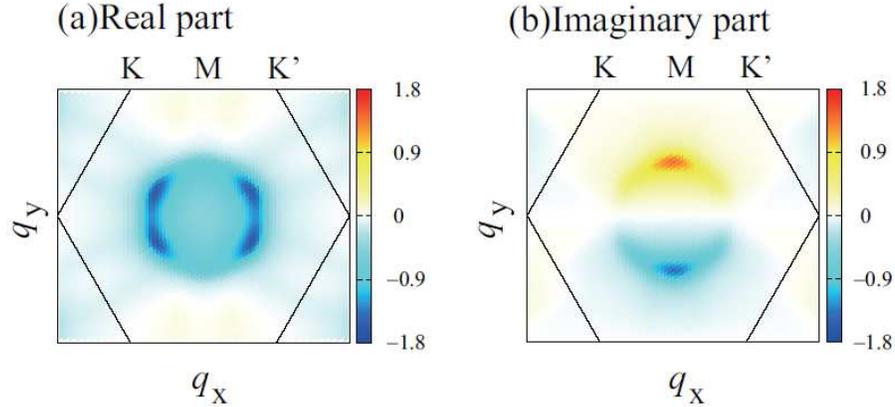}
 \caption{The longitudinal spin susceptibility 
$\chi^{zz}_{AB}(\bm{q},i\omega_m=0)$ 
for $\lambda_{\mathrm{SO}}/t=0.1$, and $U/t=2.75$: 
(a)Real part and  (b)Imaginary part. 
}
\label{LSusceptibilityAB}
\end{figure}

To understand the pairing mechanism, we show 
both longitudinal and transverse spin susceptibilities in the 
following. 
\begin{figure}[htb]
 \centering
 \includegraphics[width=6cm,clip]{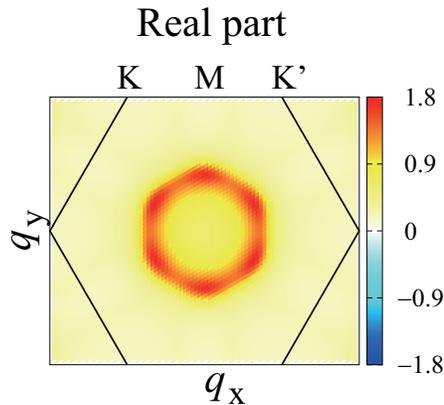}
 \caption{Real part of the longitudinal spin susceptibility 
$\chi^{zz}_{AA}(\bm{q},i\omega_m=0)$ for 
$\lambda_{\mathrm{SO}}/t=0.1$ and  $U/t=2.75$.  }
\label{LSusceptibilityAA}
\end{figure}
The longitudinal spin susceptibility 
$\chi^{zz}_{AA}(\bm{q},i\omega_m=0)$ is shown in 
Fig. \ref{LSusceptibilityAA}. 
$\chi^{zz}_{AA}(\bm{q},i\omega_m=0)$ becomes a 
real number and it
has a maximum at $\bm{q}=\bm{q}_{\mathrm{c}}$, 
where $\bm{q}_{\mathrm{c}}$ corresponds to  
a momentum transfer inside Fermi pocket. 
By contrast to $\chi^{zz}_{AA}(\bm{q},i\omega_m=0)$, 
the longitudinal spin susceptibility 
$\chi^{zz}_{AB}(\bm{q},i\omega_m=0)$ 
becomes a complex number as shown in 
Fig. \ref{LSusceptibilityAB}. 
It satisfies $\chi^{zz}_{AB}(\bm{q},i\omega_m=0)=
[\chi^{zz}_{BA}(\bm{q},i\omega_m=0)]^{*}$. 
The real part of $\chi^{zz}_{AB}(\bm{q},i\omega_m=0)$
becomes negative. 
This means that effective interaction is attractive 
one between $A$ and $B$ sublattice. 
Imaginary part of $\chi^{zz}_{AB}(\bm{q},i\omega_m=0)$ is an 
odd-function of $\bm{q}$. Then, it does not contribute 
to the actual integral kernel of the 
$\acute{\mathrm{E}}$liashberg equation. 

\begin{figure}[htbp]
 \centering
 \includegraphics[width=7cm,clip]{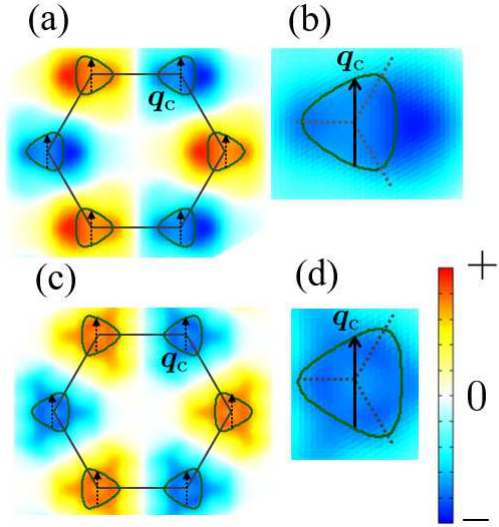}
 \caption{Schematic illustrations of the momentum transfer 
by the pair scattering at  $\bm{q}=\bm{q}_{\mathrm{c}}$. 
(a)Imaginary part of the 
energy gap function for ESE pairing 
has a sign change by pair scattering.  
(b)Enlarged view of (a). 
(c)Read part of the energy gap function for ETO $f$-wave pairing does not have a sign change 
by pair scattering.
(d)Enlarged view of (c). }
\label{momentumtransfer}
\end{figure}

The schematic image of the pair scattering 
by $\bm{q}_{\mathrm{c}}$ is shown in Fig. \ref{momentumtransfer}. 
It is noted that this pair scattering occurs 
within each disconnected Fermi surface. 
In both cases with ESE and ETO pairings, 
there is no sign change by the pair scattering 
$\bm{q}_{\mathrm{c}}$. 
However, the reasons of the absence of sign change 
are different each other.
Singlet and triplet channels of the effective interaction
originated from longitudinal susceptibility is given by
$\chi^{zz}_{\alpha\beta}(\bm{q},i\omega_{m})U^2/2$ and
$-\chi^{zz}_{\alpha\beta}(\bm{q},i\omega_{m})U^2/2$, respectively.
In the present ESE pairing, dominant pair is formed
between $A$ and $B$ sites.
In this case, $\chi^{zz}_{AB}(\bm{q},i\omega_{m})$
mainly contributes to the effective interaction.
Since the real part of 
$\chi^{zz}_{AB}(\bm{q},i\omega_{m})$ is negative,
effective interaction becomes negative, $i.e.$, attractive interaction.
On the other hands, for ETO pairing,
dominant pair is formed between the same sublattice sites.
Then, $\chi^{zz}_{AA}(\bm{q},i\omega_{m})$
mainly contributes to the effective interaction.
Though the real part of 
$\chi^{zz}_{AA}(\bm{q},i\omega_{m})$ is positive,
effective interaction becomes negative.
This is because the coefficient $-1/2$ for triplet channel
gives additional sign.
Then, the effective interaction for ETO channel also becomes attractive.
As a result, the gap function without sign change on the Fermi surface is favorable as shown in Fig. \ref{momentumtransfer}.

In order to clarify how spin 
fluctuation is influenced by $\lambda_{\mathrm{SO}}$, 
we study spin susceptibility at 
$\bm{q}=\bm{q}_{\mathrm{c}}$ and $\bm{q}=0$  
as a function of $\lambda_{\mathrm{SO}}$.
At $\lambda_{\mathrm{SO}}=0$, 
$2\chi^{zz}_{\alpha\beta}=\chi^{+-}_{\alpha\beta}$ is satisfied due to the spin- rotational symmetry.
However, this relation is broken in the presence of 
$\lambda_{\mathrm{SO}}$. 
\begin{figure}[htbp]
 \centering
 \includegraphics[width=7cm,clip]{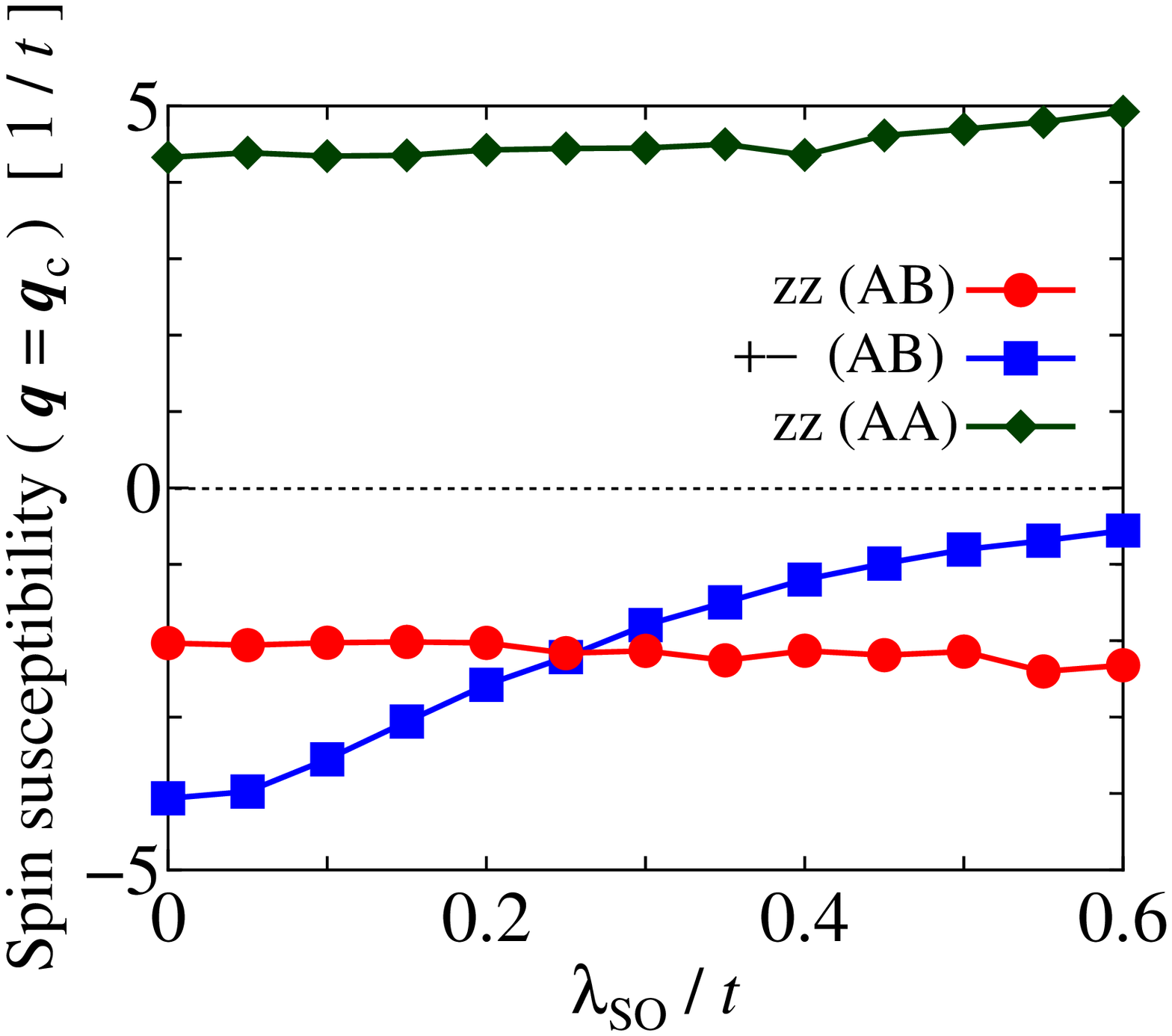}
 \caption{Longitudinal and transverse spin susceptibilities are plotted 
as a function of $\lambda_{\mathrm{SO}}$ for 
 $U=0.98U_{\mathrm{c}}$ and $\bm{q}=\bm{q}_{\mathrm{c}}$. 
$zz$
($\alpha \beta$) and $+-$($\alpha \beta$) denote 
longitudinal spin susceptibility 
$\mathrm{Re}[\chi^{zz}_{\alpha\beta}(\bm{q},i\omega_m=0)]$
and  transverse spin susceptibility 
$\mathrm{Re}[\chi^{+-}_{\alpha\beta}(\bm{q},i\omega_m=0)]$.
}
\label{Susceptibilityq=qc}
\end{figure}
First, we show the case with $\bm{q}=\bm{q}_{\mathrm{c}}$ (Fig.  \ref{Susceptibilityq=qc}). 
The magnitude of 
$\mathrm{Re}[\chi^{+-}_{AB}(\bm{q}=\bm{q}_{\mathrm{c}},i\omega_m=0)]$ is 
greatly suppressed with the 
increase of $\lambda_{\mathrm{SO}}$. 
On the other hand, 
the magnitude of 
$\mathrm{Re}[\chi^{zz}_{AA}(\bm{q}=\bm{q}_{\mathrm{c}},i\omega_m=0)]$ 
and $\mathrm{Re}[\chi^{zz}_{AB}(\bm{q}=\bm{q}_{\mathrm{c}},i\omega_m=0)]$
are little bit enhanced. 
\begin{figure}[htbp]
 \centering
 \includegraphics[width=7cm,clip]{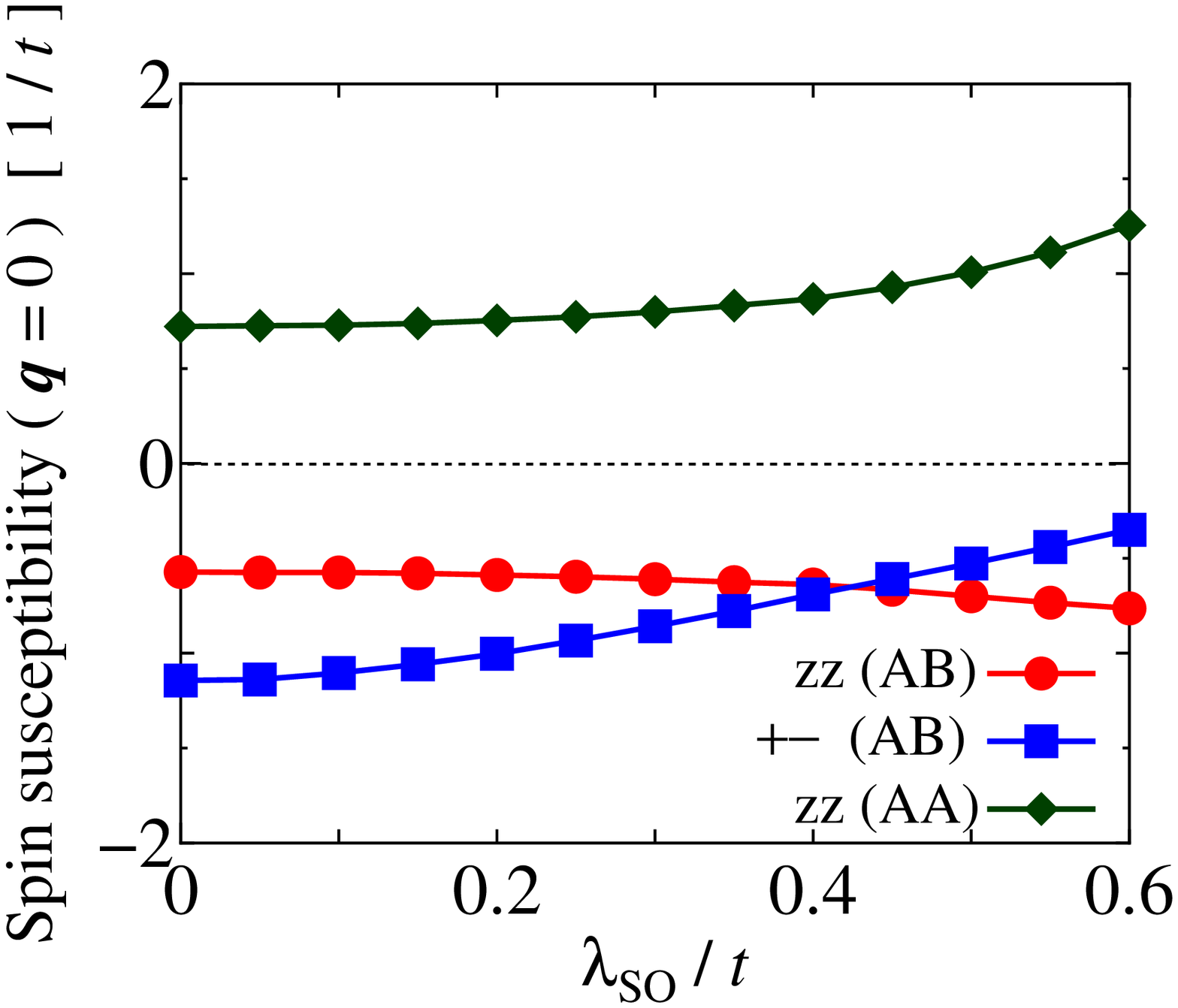}
 \caption{Longitudinal and transverse spin susceptibilities are plotted 
as a function of $\lambda_{\mathrm{SO}}$ for 
 $U=0.98U_{\mathrm{c}}$ and $\bm{q}=0$. 
$zz$
($\alpha \beta$) and $+-$($\alpha \beta$) denote 
longitudinal spin susceptibility 
$\chi^{zz}_{\alpha\beta}(\bm{q},i\omega_m=0)$
and  transverse spin suceptibilitibility 
$\chi^{+-}_{\alpha\beta}(\bm{q},i\omega_m=0)$.}
\label{Susceptibilityq=0}
\end{figure}
Next, we show the case with $\bm{q}=0$ (Fig. \ref{Susceptibilityq=0}). 
The magnitude of 
$\chi^{+-}_{AB}(\bm{q}=0,i\omega_m=0)$ is 
suppressed with the 
increase of $\lambda_{\mathrm{SO}}$ similar to the case of 
$\bm{q}=\bm{q}_{\mathrm{c}}$. 
On the other hand, 
the magnitude of $\chi^{zz}_{AA}(\bm{q}=0,i\omega_m=0)$ 
and $\chi^{zz}_{AB}(\bm{q}=\bm{q}_{\mathrm{c}},i\omega_m=0)$
are enhanced. 
The degree of enhancement of 
$\chi^{zz}_{AA}(\bm{q}=0,i\omega_m=0)$ is 
greater than that of 
$\chi^{zz}_{AB}(\bm{q}=0,i\omega_m=0)$. 

\begin{figure}[htbp]
 \centering
 \includegraphics[width=7cm,clip]{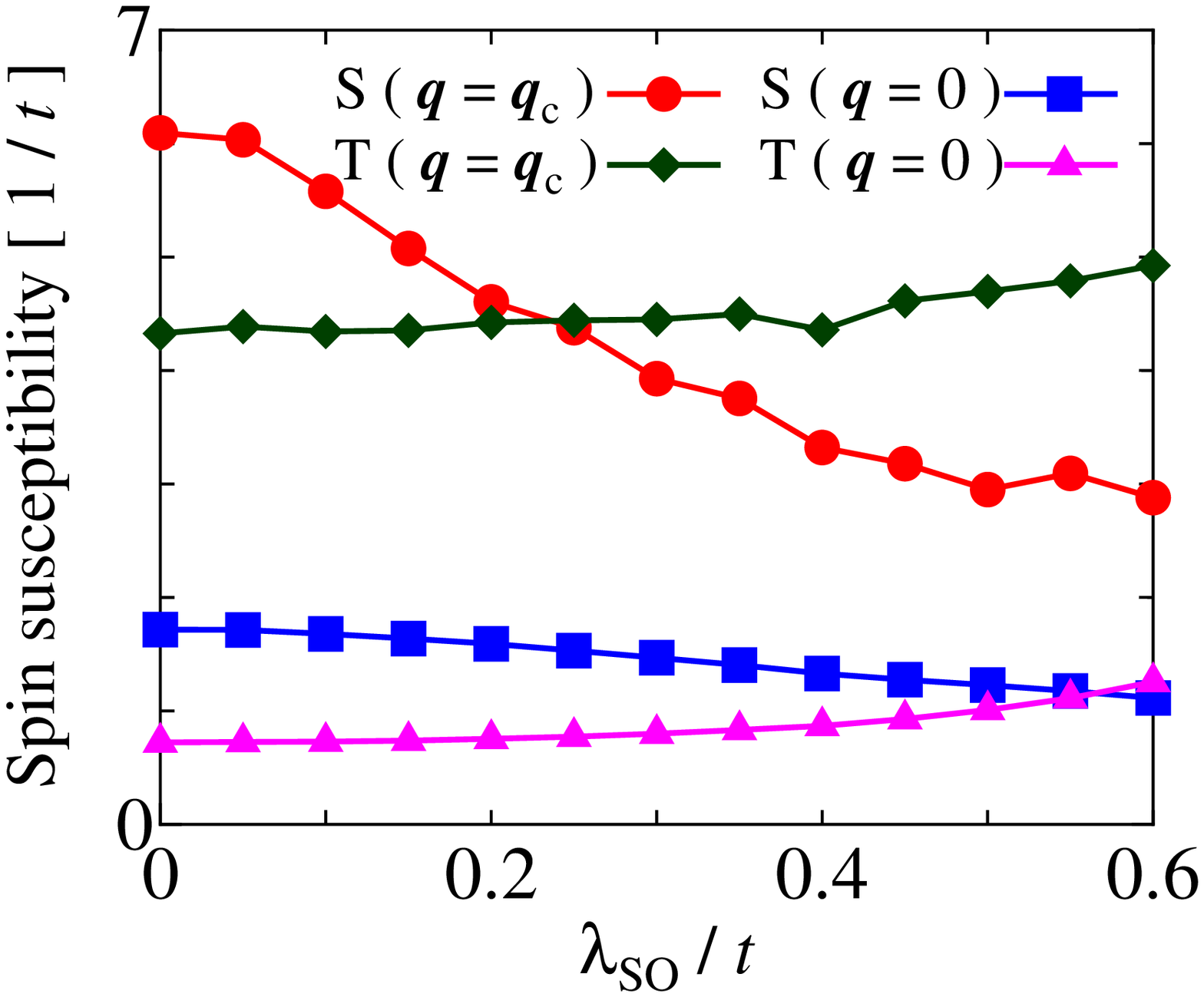}
 \caption{The absolute value of spin susceptibilities are plotted as a function of $\lambda_{\mathrm{SO}}$ for
 $U=0.98U_{\mathrm{c}}$ with  $\bm{q}=\bm{q}_{\mathrm{c}}$ and 
$\bm{q}=0$. 
$S(\bm{q}=\bm{q}_{\mathrm{c}})= 
| 
\mathrm{Re}[\chi^{zz}_{AB}(\bm{q}_{\mathrm{c}},i\omega_m=0)
+\chi^{+-}_{AB}(\bm{q}_{\mathrm{c}},i\omega_m=0)]| $,  
$S(\bm{q}=0)= |
\mathrm{Re}[
\chi^{zz}_{AB}(0,i\omega_m=0)
+\chi^{+-}_{AB}(0,i\omega_m=0)]|$, 
$T(\bm{q}=\bm{q}_{\mathrm{c}}) =|
\mathrm{Re}[\chi^{zz}_{AA}(\bm{q}_{\mathrm{c}},i\omega_m=0)]|$, 
and 
$T(\bm{q}=0) =
|\mathrm{Re}[\chi^{zz}_{AA}(0,i\omega_m=0)]|.$}
\label{interaction}
\end{figure}
In Fig. \ref{interaction}, to see the 
strength of pairing interaction, we plot 
\begin{align}
S(\bm{q}=\bm{q}_{\mathrm{c}})&= 
| 
\mathrm{Re}[\chi^{zz}_{AB}(\bm{q}_{\mathrm{c}},i\omega_m=0)
+\chi^{+-}_{AB}(\bm{q}_{\mathrm{c}},i\omega_m=0)]|, \\ 
S(\bm{q}=0)&= |
\mathrm{Re}[
\chi^{zz}_{AB}(0,i\omega_m=0)
+\chi^{+-}_{AB}(0,i\omega_m=0)]|, \\
T(\bm{q}=\bm{q}_{\mathrm{c}}) &=|
\mathrm{Re}[\chi^{zz}_{AA}(\bm{q}_{\mathrm{c}},i\omega_m=0)]|,\\ 
T(\bm{q}=0) &=
|\mathrm{Re}[\chi^{zz}_{AA}(0,i\omega_m=0)]|.
\end{align}
Here, $S(\bm{q})$
and $T(\bm{q})$
express the spin fluctuation 
which contributes to ESE pairing and ETO pairing, respectively. 
As seen in  Fig. \ref{interaction}, the 
magnitude of $S(\bm{q}=\bm{q}_{\mathrm{c}})$ is 
strongly suppressed by spin-orbit coupling 
$\lambda_{\mathrm{SO}}$.  
The magnitude of $S(\bm{q}=0)$ is also reduced. 
This is the reason that ESE pairing becomes  destabilized 
with $\lambda_{\mathrm{SO}}$.
On the other hand, $T(\bm{q}=\bm{q}_{\mathrm{c}})$ and 
$T(\bm{q}=0)$ are enhanced with $\lambda_{\mathrm{SO}}$. 
This is the reason why ETO pairing becomes  
dominant with $\lambda_{\mathrm{SO}}$.

%
Summarizing these results, ESE pairing is 
induced by transverse spin fluctuation at $\bm{q}=\bm{q}_{\mathrm{c}}$. 
On the other hand, ETO $f$-wave pairing is 
supported by longitudinal  spin fluctuation at 
$\bm{q}=\bm{q}_{\mathrm{c}}$ and $\bm{q}=0$. 
Since the transverse spin fluctuation is strongly 
suppressed by spin-orbit coupling, ETO $f$-wave pairing 
becomes dominant for large $\lambda_{\mathrm{SO}}$. 

\section{Summary}
In this paper, we have studied 
possible pairing symmetries of a doped Kane-Mele model on the honeycomb
lattice with on-site Coulomb interaction. 
We have clarified the pairing instability of Cooper pair 
by the linearized $\acute{\mathrm{E}}$liashberg 
equation within RPA. 
When the magnitude of the spin-orbit coupling is weak, 
ESE pairing becomes dominant one.
Since Cooper pair is formed between $A$ and $B$ sites in this pairing, 
it has a complicated momentum dependence. 
In our choice of the gauge, real part has a $d$-wave symmetry 
while imaginary part has a $f$-wave like symmetry.
This $f$-wave like pairing 
does not contradict even-parity pairing 
because it has 
a sign change with the exchange of the index $A$ and $B$. 
At the same time, 
OTE pairing with $S_{z}=0$  
also mixes as a subdominant component of a 
solution of the $\acute{\mathrm{E}}$liashberg equation. 
It is triggered by the 
intrinsic spin-orbit coupling 
which does not flip the spin. 
By contrast to ESE dominant case, 
odd-frequency subdominant pair never appears 
since spin-triplet $f$-wave pair is composed of two electrons with 
equal spin. 
With the increase of the magnitude of spin-orbit coupling, 
we have clarified that the spin-triplet $f$-wave 
pairing becomes dominant. 
This is because the transverse spin susceptibility 
is suppressed by spin-orbit coupling and 
the resulting effective interaction for ESE channel 
is weakened. \par
In this paper, we have focused on the pairing mechanism of 
doped Kane-Mele model and found the instability of 
unconventional superconductivity. 
Nowadays, it is known  
that both even  and odd-parity pairings 
discussed in this paper have surface Andreev bound states (SABS)
which are protected by topological invariants \cite{index,Tanaka12}. 
It is interesting to calculate SABS and tunneling spectroscopy 
via Andreev bound state \cite{TK95,Tanuma2001,Tanuma2002} 
in order to distinguish spin-triplet odd-parity pairing from 
spin-singlet even-parity one. 
Especially charge transport in 
diffusive normal metal / spin-triplet odd-parity superconductor 
is interesting since we have obtained 
anomalous proximity effect by odd-frequency pairing and Majorana fermion
in diffusive normal metal / spin-triplet $p$-wave superconductor 
junctions \cite{Proximityp,Proximityp2,odd1,Asano2013}. 

%
%
\begin{acknowledgments}
This work was supported by a Grant-in-Aid for Scientific 
Research on Innovative Areas Topological Material
Science (Grant Nos. 15H05853 and 15H05851), a Grant-in-Aid for Scientific 
Research B (Grant No. 15H03686), 
a Grant-in-Aid for Challenging Exploratory Research (Grant No.
15K13498) from the Ministry of Education, Culture,
Sports, Science, and Technology, Japan (MEXT). 
\end{acknowledgments}
%
%
\bibliographystyle{jpsj}
\bibliography{fukaya7}

\end{document}